\documentclass[final]{siamltexmm1}

\usepackage{multirow}
\usepackage{moreverb}
\usepackage{graphicx}
\RequirePackage{subfigure}[1995/03/06]
\usepackage{amssymb}
\usepackage{amsmath}
\usepackage{color}

\newcommand{\HL}{\color{red}}

\newcommand{\XX}{\textbf{XX}}

\title{A DYNAMIC BI-ORTHOGONALITY BASED APPROACH FOR UNCERTAINTY QUANTIFICATION OF STOCHASTIC \\ SYSTEMS WITH DISCONTINUITIES\thanks{This work was supported by Basic Science Research Program through the National Research Foundation of Korea (NRF) funded by the Ministry of Education, Science and Technology (2010-0025484)}}

\author{Piyush M. Tagade\thanks{Post-doctoral Research Fellow, Division of Aerospace Engineering, Korea Advanced Institute of Science and Technology, Republic of Korea. (\email{piyush.tagade@kaist.ac.kr}).}
\and Han-Lim Choi\thanks{Assistant Professor, Division of Aerospace Engineering, Korea Advanced Institute of Science and Technology, Republic of Korea. (\email{hanlimc@kaist.ac.kr}). Questions, comments, or corrections to this document may be directed to that email address.}}

\begin{document}
\maketitle
\newcommand{\slugmaster}{%
\slugger{MMedia}{xxxx}{xx}{x}{}}

\begin{abstract}
The use of spectral projection based methods for simulation of a stochastic system with discontinuous solution exhibits the Gibbs phenomenon, which is characterized by oscillations near discontinuities.
This paper investigates a dynamic bi-orthogonality based approach with appropriate post-processing for mitigating the effects of the Gibbs phenomenon.
The proposed approach uses spectral decomposition of the spatial and stochastic fields in appropriate orthogonal bases, while the dynamic orthogonality condition is used to derive the resultant closed form evolution equations.
The orthogonal decomposition of the spatial field is exploited to propose a Gegenbauer reprojection based post-processing approach, where the orthogonal bases in spatial dimension are reprojected on the Gegenbauer polynomials in the domain of analyticity.
The resultant spectral expansion in Gegenbauer series is shown to mitigate the Gibbs phenomenon.
Efficacy of the proposed method is demonstrated for simulation of a one-dimensional stochastic Burgers equation with uncertain initial condition.

\end{abstract}

\begin{keywords} Uncertainty Quantification, Gibbs Phenomenon, Stochastic Spectral Methods, Dynamically Bi-orthogonal Field Equations \end{keywords}

\begin{AMS} 74S25, 74S60, 74J40 \end{AMS}

\pagestyle{myheadings}
\thispagestyle{plain}
\markboth{PIYUSH TAGADE AND HAN-LIM CHOI}{DBFE FOR SPDEs WITH DISCONTINUITIES}

\section{Introduction}
Continuous advancements in digital technologies have resulted in ubiquitous use of computer simulators for investigation of many large-scale systems.
The simulator predictions are often uncertain due to unknown/poorly known physics, model parameters, initial and boundary conditions.
Need and importance of uncertainty quantification in simulation predictions has already been emphasized by researchers in varied fields~\cite{OreskesScience94,Mehta91,Mehta96,OberkampfRESS02}.
Subsequently, research community have invested significant efforts to develop various uncertainty quantification methodologies~(see \cite{Walters02} and references therein.).
In essence, uncertainty quantification is an estimate of joint probability distribution of system responses conditional on probabilistic specification of uncertainties in the simulator model, parameters, initial and boundary conditions etc.

\subsection{Background}
For further discussion, consider a probability space $\left(\Omega, \mathcal{F}, \mathcal{P} \right)$ where $\omega \in \Omega$ is a set of elementary events, $\mathcal{F}$ is associated $\sigma$-algebra and $\mathcal{P}$ is a probability measure defined over $\mathcal{F}$.
A generic stochastic partial differential equation (SPDE)  on $\left(\Omega, \mathcal{F}, \mathcal{P} \right)$ is given by
\begin{equation}
\frac{\partial u(x,t;\omega)}{\partial t} = \mathcal{L} \left[u(x,t;\omega);\omega \right]
\label{spde} 
\end{equation}
where $u(x,t;\omega)$ is a random variable, $t\in \mathcal{T}$ is time, $x\in \mathcal{X} \subset \mathcal{R}^d$ is a spatial dimension and $\mathcal{L}$ is an arbitrary differential operator of the form
\begin{equation}
\mathcal{L} \left[u(x,t;\omega);\omega \right] = -\frac{\partial f(u(x,t;\omega);\omega)}{\partial x},
\end{equation}
$f(u(x,t;\omega);\omega)$ being flux.
The initial condition of the system is specified using a random field $u(x,0;\omega)$, while, the boundary conditions are given by
\begin{equation}
\mathcal{B}(\beta,t;\omega) = h(\beta,t;\omega); ~ ~ ~ \beta \in \partial \mathcal{X}, ~ \omega \in \Omega,
\label{bound_cond}
\end{equation}
where $\mathcal{B}$ is a linear differential operator.
Without loss of generality, the discussion in this paper is presented assuming $u(x,t;\omega)$ to be a scalar function defined over a one-dimensional space, however, the discussion can be extended to a more generic case in straightforward manner.

Owing to generality and ease of implementation, Monte Carlo methods are widely used for solution of (\ref{spde}) \cite{rubinstein}.
However, the Monte Carlo method typically requires a large number of samples for acceptable accuracy~\cite{rubinstein,chopin_04}.
For simulation of large scale systems, computational cost may render application of Monte Carlo methods for uncertainty quantification intractable. %
Stochastic spectral projection (SSP) based methods can provide a  computationally efficient alternative to Monte Carlo methods with comparable accuracy. 
Under a generic condition of square integrability, $u(x,t;\omega)$ can be represented as a parametric field in a Hilbert space $\mathcal{H}(\mathcal{T} \times \mathcal{X} \times \mathcal{F})$~\cite{Adler}.
The SSP methods use the tensor product representation of $\mathcal{H}(\mathcal{T} \times \mathcal{X} \times \mathcal{F})$ in appropriate Hilbert spaces $\mathcal{H}_1$ and $\mathcal{H}_2$ with possible choices ~\cite{venturi_11}
\begin{eqnarray}
\mathcal{H}(\mathcal{T} \times \mathcal{X} \times \mathcal{F}) & = &  \mathcal{H}_1(\mathcal{T}  \times \mathcal{F}) \times \mathcal{H}_2(\mathcal{X}) \label{tens_prod_1} \\
~ & = & \mathcal{H}_1(\mathcal{T}) \times \mathcal{H}_2(\mathcal{X} \times \mathcal{F}) \label{tens_prod_2} \\
~ & = & \mathcal{H}_1(\mathcal{T} \times \mathcal{X}) \times \mathcal{H}_2(\mathcal{F}) \label{tens_prod_3}.
\end{eqnarray}
A Karhunnen-Loeve expansion~\cite{Ghanem} uses the tensor product (\ref{tens_prod_1}), where $u(x,t;\omega)$ is spectrally represented in terms of orthogonal basis in $\mathcal{H}_2(\mathcal{X})$ with temporally varying expansion coefficients in $\mathcal{H}_1(\mathcal{T}  \times \mathcal{F})$.
A generalized Polynomial Chaos (gPC) expansion~\cite{XiuJCP03} is an example of tensor product (\ref{tens_prod_3}) with orthogonal basis in $\mathcal{H}_2(\mathcal{F})$ and the associated expansion coefficients in $\mathcal{H}_1(\mathcal{T} \times \mathcal{X})$.
This paper investigates a bi-orthogonal method, which uses the gPC basis in $\mathcal{H}_2(\mathcal{F})$, while, dynamically orthogonal eigenfunction basis is used in $\mathcal{H}_1(\mathcal{T} \times \mathcal{X})$.

One of the earliest exposition of a SSP-based method is the Homogeneous Chaos theory introduced by Wiener~\cite{WienerAJM38,Wiener}, where random variables are expanded in terms of orthogonal Hermite polynomials.
The expansion converges in mean square sense for any nonlinear $L^2$ functional~\cite{CameronAM47}.
Meecham and Jeng~\cite{MeechamJFM68} used Homogeneous Chaos for turbulent modeling, however it was found that the convergence of chaos expansion is slow for turbulent field~\cite{OrszagPF67,ChorinJFM74}.
Ghanem and Spanos~\cite{Ghanem91,GhanemPhyD99,Ghanem} have proposed Polynomial Chaos (PC) method based on Homogeneous Chaos theory for uncertainty propagation in structural mechanics simulations.
Subsequently, the method has been widely used for uncertainty propagation in fluid flow simulations~\cite{jcp_stoch_proj_2001,Lucor03,Mathelin04,Narayanan04}.
Present state of the art for SSP methods is based on the Galerkin projection based generalized Polynomial Chaos method introduced by Xiu and Karniadakis~\cite{Xiu_SJSC_2004,XiuJCP03}, which uses the representation of form (\ref{tens_prod_1}).
Using separability, $\mathcal{H}_1(\mathcal{F})$ is decomposed in orthogonal polynomial chaos basis $\psi_k(\omega)$, where $\psi_k(\omega)$ belong to the Askey scheme of orthogonal polynomials~\cite{Koekoek} with weight function given by probability distribution function of an uncertain input.
Thus, using gPC, the dynamical response $u(x,t;\omega)$ is spectrally represented as
\begin{equation}
u(x,t;\omega) = \sum^{P}_{i=1} \hat{u}_i(x,t) \psi_i(\omega),
\label{gpc}
\end{equation}
where $\hat{u}_i(x,t)$ are dynamical expansion coefficients.
See Mathelin et al.~\cite{Mathelin_NA_05} and Najm~\cite{Najm_arfm_09} for extensive review of the research work in this field.
\newtheorem{rmkequal}{Remark}
\begin{rmkequal}
Throughout this paper, the operator $`='$ is used interchangeably to represent `equal to' and `approximately equal to'.
Note that all the instances of $`='$ in spectral representation defines the `approximately equal to' operator.
\end{rmkequal}

\subsection{Motivation}
Research work presented in this paper is motivated by the need to quantify uncertainty in simulation of systems with discontinuous solutions.
In particular, the paper deals with stochastic systems defined using the hyperbolic conservation laws that invariably results in discontinuous solution irrespective of the initial conditions.
Simulation of such system have attracted significant interest by research community and literature is rich with methods for resolution of discontinuities in deterministic simulations~(see \cite{LeVeque} and references therein for review of the methods).

The use of spectral methods to hyperbolic SPDEs is known to pose following significant numerical challanges~\cite{najm_11}
\begin{enumerate}
\item numerical solution using finite difference schemes lead to spurious oscillations in expansion coefficients and/or orthogonal basis
\item spectral expansion of the solution of hyperbolic SPDEs lead to Gibbs phenomenon as discontinuities develop in solution~\cite{najm_11}.
The Gibbs phenomenon is characterized by~\cite{Hewitt_79,gottlieb_97} a) slow convergence of spectral approximation of the solution at points away from the discontinuity; and b) $\mathcal{O}(1)$ oscillations are observed in the solution near discontinuities that do not decrease with increasing $N$.
\end{enumerate}

There are recent research efforts that focus on addressing the issue of Gibbs phenomenon in gPC settings \cite{najm_11}.
Wan and Karniadakis \cite{wan_05} have proposed a multi-element gPC (ME-gPC) method to mitigate the effect of Gibbs phenomenon (also see Lin et al. \cite{lin_06}).
The ME-gPC method uses domain decomposition of the stochastic space with locally defined gPC basis, while the forward problem is solved for each subdomain.
However, the computational cost of the ME-gPC method rises quickly as the number of subdomains increases.
Poette et al. \cite{PoetteJCP09} have proposed an entropy based intrusive polynomial moment method for uncertainty propagation in non-linear systems with discontinuous solutions.
The method reformulates the SPDE in terms of appropriately selected entropic variables that are bijection of the uncertain parameters, while, the gPC expansion of the entropic variables is defined through the Galerkin projection of the bijection.
The method is implemented in two steps: in the first step, the resultant stochastic Galerkin system is discretized using a finite volume approach, and upwinded Row scheme is used for numerical solution.
In the second step, gPC expansion of the entropic variables is calculated so that the entropy of the system is minimized.
The method is demonstrated for solution of the stochastic Burgers equation.
The method is computationally efficient than the ME-gPC, however, the requirement for minimization of the entropy at each time step results in the computational cost higher than the gPC method.
Tryoen et al. \cite{tryoyen_10} uses Roe-type solver for intrusive Galerkin projection of uncertain hyperbolic systems.
The solver uses Dubois and Mehlman entropy corrector that avoid certain entropy violating solutions.
Sargsyan et al. \cite{sargsyan_12} have proposed a two step method for uncertainty quantification of systems with discontinuous response.
In the first step, a limited number of simulation runs are used to infer the shock location using the Bayesian inference.
In the second step, localized gPC basis is defined in the region of analiticity, while intrusive Galerkin projection is used to propagate the uncertainty.
In an alternate non gPC setting, Chantrasmi et al. \cite{chantr_09} have proposed a Pade-Legendre interpolant based approach, where, simulation runs are obtained at predefined Gauss-Legendre quadrature nodes and the resultant discontinuous system response is reconstructed using the Pade interpolation.
Despite of these recent progresses, resolution of Gibbs phenomenon in application of SSP methods to stochastic hyperbolic systems is an area of current research interest~\cite{sargsyan_12}.

\subsection{Proposed Method}
Existence and resolution of Gibbs phenomenon for deterministic functions is extensively investigated in the literature~\cite{Hewitt_79,gottlieb_97,gottlieb_11}.
Spectral expansion of a function using global orthogonal basis is contaminated by the presence of discontinuities.
However, expansion coefficients contain enough information to recover the function with high accuracy using appropriate post-processing.
Gibbs phenomenon can be completely resolved by reprojecting the partial sum on Gibbs complementary basis~\cite{gottlieb_97}.
Gottlieb et al.~\cite{gottlieb_1,gottlieb_4,gottlieb_97,gottlieb_11} have shown that exponentially convergent approximation can be obtained at point values in subinterval of analyticity by reprojecting the partial sum on orthogonal Gegenbauer polynomial basis.
The post-processing method can be extended to stochastic functions, provided, the spectral expansion is available in terms of orthogonal basis in spatial dimensions.

Note that the gPC expansion (\ref{gpc}) uses orthogonal basis in $\mathcal{H}_2(\mathcal{F})$  while corresponding coefficients $\hat{u}_i(x,t)$ are functions in $\mathcal{H}_1(\mathcal{T} \times \mathcal{X})$, though, the discontinuity reside in spatial dimension.
Thus, enough information is not available in $\hat{u}_i(x,t)$ to resolve Gibbs phenomenon using Gegenbaeur reprojection method.
Resolution of Gibbs phenomenon proposed in this paper is based on the observation that exponential accuracy can be recovered if $u(x,t;\omega)$ is expanded in terms of orthogonal bases in  \emph{spatial}  dimension, and subsequently reprojected on an approximate Gibbs complementary basis.

Tagade and Choi \cite{TagadeIDETC12, TagadeCSDA12} have proposed a dynamic bi-orthogonal field equations (DBFE) method for solution of SPDEs in the context of Bayesian calibration.
This paper proposes a method for solution of hyperbolic SPDEs using DBFE, that exploits orthogonal decomposition of the spatial dimension to develop a post-processing approach for mitigation of the Gibbs phenomenon.   
Consider a generic Karhunnen-Loeve expansion of $u(x,t;\omega)$
\begin{equation}
u(x,t;\omega) = \overline{u}(x,t) + \sum^{N}_{i=1} Y_i(t;\omega) u_i(x,t)
\label{kl_expn}
\end{equation}
where $\overline{u}(x,t)$ is the mean, $u_i(x,t)$ are eigenfunctions which form complete orthogonal basis in $\mathcal{H}_1(\mathcal{X})$ at a given time step $t$, while $Y_i(t;\omega)$ are independent zero-mean random variables.
Using polynomial chaos expansion of $Y_i(t;\omega)$ in (\ref{kl_expn}), bi-orthogonal expansion of $u(x,t;\omega)$ is given by
\begin{equation}
u(x,t;\omega) = \overline{u}(x,t) + \sum^{N}_{i=1} \sum^{P}_{p=1} \hat{Y}^i_p(t) \psi_p(\omega) u_i(x,t).
\label{kl_expn_dbfe}
\end{equation}
A Dynamic Orthogonality (DO) condition \cite{pierre_09} is used to derive closed form solution of evolution equations for $\overline{u}(x,t)$, $u_i(x,t)$ and $\hat{Y}^i_p(t)$.
However, the resultant approximation of $u(x,t;\omega)$ in (\ref{kl_expn_dbfe}) using the solution of $\overline{u}(x,t)$, $u_i(x,t)$ and $\hat{Y}^i_p(t)$ exhibits the Gibbs phenomenon.
A Gegenbauer reprojection based post-processing method is proposed to mitigate the effect of Gibbs phenomenon by reprojecting the eigenfunctions $u_i(x,t)$ on the Gegenbauer basis.   
The proposed method is demonstrated for solution of a stochastic one-dimensional Burgers equation with uncertain initial conditions.

The rest of the paper is organized as follows. In section 2, mathematical formulation of the method is provided.
Proposed post-processing method for Resolution of the Gibbs phenomenon using dynamic bi-orthogonality based method is discussed in section 3.
Section 4 provides numerical results for Burgers equation to demonstrate efficacy of the proposed method.
Finally, paper is summarized and concluded in section 5.

\section{Dynamically Bi-orthogonal Field Equations (DBFE)}

Definition of the following operators will be used in the remainder of the paper.
\newtheorem{inpdef}{Definition}
\begin{inpdef}
Inner product on $\mathcal{H}_1(\mathcal{T} \times \mathcal{X})$ for a given $t$ is defined as
\begin{equation}
\left< u(x,t;\omega),v(x,t;\omega)\right>_{\mathcal{X}} = \int_{\mathcal{X}} u(x,t;\omega) v(x,t;\omega) dx .
\label{inn_h1}
\end{equation}
Similarly, inner product on $\mathcal{H}_2(\mathcal{F})$ is defined as
\begin{equation}
\left\langle u(x,t;\omega),v(x,t;\omega)\right\rangle_{\Omega} = \int_{\Omega} u(x,t;\omega) v(x,t;\omega) d\mathcal{P}(\omega) .
\label{inn_h2}
\end{equation}
The expectation operator on $\mathcal{H}_2(\mathcal{F})$ is defined as
\begin{equation}
\overline{u}(x,t) = {E}^{\omega} [u(x,t;\omega)] = \int_{\Omega} u(x,t;\omega) d\mathcal{P}(\omega).
\label{exp_value}
\end{equation}
Further using definition of the expectation operator, the covariance is defined as
\begin{equation}
C_{u,v} = {E}^{\omega} [u(x,t;\omega) v(x,t;\omega)].
\label{cov_oper}
\end{equation}
\end{inpdef}

The DBFE method uses the  bi-orthogonal expansion (\ref{kl_expn_dbfe}) of $u(x,t;\omega)$ in (\ref{spde}).
However, independent evolution equations for the mean $\overline{u}(x,t)$, the eigenfunctions $u_i(x,t)$ and the corresponding coefficients $Y^i_p(t)$ can not be obtained concurrently using the SPDE in (\ref{spde}).
Well posed evolution equations for the unknown quantities can de derived by imposing the additional dynamic orthogonality (DO) condition~\cite{pierre_09}.
The DO condition is specified as \cite{pierre_09}
\begin{equation}
\left\langle u_i(x,t),\frac{\partial u_j(x,t)}{\partial t} \right\rangle_{\mathcal{X}} = 0, ~~\forall i,~j=1,...,N,
\label{do_cond}
\end{equation}
where $N$ is the number of eigenfunctions retained in the expansion (\ref{kl_expn_dbfe}).
Note that the DO condition constrains the dynamic evolution of the eigenfunction basis such that the orthonormality of the eigenfield is retained at all time steps \cite{pierre_09}.

Using the DO condition and bi-orthogonal expansion of $u(x,t;\omega)$,  SPDE (\ref{spde}) can be reformulated into a set of $N+1$ PDEs and $N\times P$ ODEs as follows.
\newtheorem{prop2}{Proposition}
\begin{prop2}
Using the DO condition, dynamic evolution equations of $\overline{u}(x,t)$, $u_i(x,t)$ and $\hat{Y}^i_p(t)$ are given by
\begin{eqnarray}
\frac{\partial \overline{u}(x,t)}{\partial t} &  = & E^{\omega}\left[\mathcal{L}[u(x,t,;\omega);\omega]\right] \label{mean_ev} \\
\frac{\partial u_i(x,t)}{\partial t} & = & \sum^{N}_{j=1} C^{-1}_{Y_i Y_j} \Big\{E^{ \omega}[\mathcal{L}[u(x,t,\omega);\omega] Y_j(t;\omega)] \notag \\
~ & ~ & \qquad - \sum_{k=1}^{N} \left\langle E^{\omega}[\mathcal{L}[u(x,t,\omega);\omega] Y_j(t;\omega)],u_k(x,t)\right\rangle_{\mathcal{X}} \Big\} \label{efun_ev} \\
\frac{d Y^i_p(t)}{dt} & = & \frac{1}{\left\langle \psi^2_p(\omega) \right\rangle_{\Omega}}\times \notag \\  ~& ~& \left\langle \left\langle \mathcal{L}[u(x,t;\omega);\omega] - E^{\omega} [\mathcal{L}[u(x,t;\omega);\omega]], u_i(x,t)\right\rangle_{\mathcal{X}}, \psi_p(\omega)\right\rangle_{\Omega} \label{eyp_ev}.
\end{eqnarray}
\end{prop2}
\begin{proof}
Proof of (\ref{mean_ev}) and (\ref{efun_ev}) is given by Sapsis and Lermusiaux~\cite{pierre_09}, which is briefly presented here for completeness, while, (\ref{eyp_ev}) is derived here by introducing the bi-orthogonality.
\paragraph{{Proof of (\ref{mean_ev}) and (\ref{efun_ev})}}
Use a generic KL expansion (\ref{kl_expn}) in (\ref{spde}) to obtain
\begin{equation}
\frac{\partial \overline{u}(x,t)}{\partial t} + \sum_{i=1}^{N} u_i(x,t) \frac{d Y_i(t;\omega)}{d t} + \sum_{i=1}^{N} Y_i(t;\omega) \frac{\partial u_i(x,t)}{\partial t} = \mathcal{L}\left[u(x,t;\omega);\omega\right].
\label{spde_kl}
\end{equation}
Application of the expectation operator to (\ref{spde_kl}) gives (\ref{mean_ev}).

Multiply (\ref{spde_kl}) by $Y_j(t;\omega)$ and apply the expectation operator to obtain
\begin{equation}
\sum^{N}_{i=1} C_{\frac{d Y_i(t)}{dt}Y_j(t)} u_i(x,t) + \sum^{N}_{i=1} C_{Y_i(t)Y_j(t)} \frac{\partial u_i(x,t)}{\partial t} = E^{\omega}\left[\mathcal{L}\left[u(x,t;\omega);\omega\right] Y_j(t;\omega)\right].
\label{cov_dyidt}
\end{equation}
Multiply (\ref{cov_dyidt}) by $u_k(x,t)$, take inner product and apply the DO condition to get
\begin{equation}
 C_{\frac{d Y_k(t)}{dt}Y_j(t)} = \left\langle E^{\omega}\left[\mathcal{L}\left[u(x,t;\omega);\omega\right] Y_j(t;\omega)\right], u_k(x,t)\right\rangle_{\mathcal{X}} .
\label{cov_dyidtuk}
\end{equation}
Use (\ref{cov_dyidtuk}) in (\ref{cov_dyidt}) to obtain
\begin{eqnarray}
\sum^{N}_{i=1} C_{Y_i(t) Y_j(t)} \frac{\partial u_i(x,t)}{\partial t} & = & E^{\omega}\left[\mathcal{L}[ u(x,t,\omega);\omega] Y_j(t;\omega) \right] \\
~ & ~ & - \sum^{N}_{k=1} \left\langle E^{\omega}\left[ \mathcal{L}[u(x,t,\omega);\omega] Y_j(t;\omega)\right],u_k(x,t) \right\rangle_{\mathcal{X}} u_k(x,t) ,
\end{eqnarray}
which can be written in the matrix as
\begin{equation}
{\bf U} = \Sigma^{-1} {\bf D},
\end{equation}
where $\Sigma$ is the covariance matrix with $ij^{th}$ element $\Sigma_{ij} = C_{Y_i Y_j}$.

Multiply both sides of (\ref{spde_kl}) by $u_j(x,t)$, take inner product and use the DO condition and orthonormality of eigenfunctions to get
\begin{equation}
\left \langle \frac{\partial \overline{u}(x,t)}{\partial t},  u_j(x,t)\right \rangle_{\mathcal{X}} + \frac{d Y_j(t;\omega)}{d t} = \left \langle \mathcal{L}\left[u(x,t;\omega);\omega\right], u_j(x,t)\right \rangle_{\mathcal{X}},
\label{inn_prod_uij_1}
\end{equation}
which on application of expectation operator gives
\begin{equation}
\left \langle \frac{\partial \overline{u}(x,t)}{\partial t}, u_j(x,t)\right \rangle_{\mathcal{X}} = E^{\omega}\left[\left \langle \mathcal{L}\left[u(x,t;\omega);\omega\right], u_j(x,t)\right \rangle_{\mathcal{X}} \right].
\label{inn_prod_uij_2}
\end{equation}
Using (\ref{inn_prod_uij_2}) in (\ref{inn_prod_uij_1}) gives
\begin{equation}
\frac{d Y_i(t;\omega)}{d t} = \left \langle \left[ \mathcal{L}\left[(x,t;\omega);\omega\right] -  E^{\omega}\left[\mathcal{L}\left[u(x,t;\omega);\omega\right]\right]\right], u_i(x,t)\right \rangle_{\mathcal{X}}.
\label{gov_eq2}
\end{equation}

\paragraph{Proof of (\ref{eyp_ev})}
Basic conditions of KL expansion ensures that $Y_i(t;\omega)$ are square integrable random variables, thus, according to Cameron and Martin theorem~\cite{CameronAM47}, $Y_i(t;\omega)$ can be approximated to arbitrary accuracy using spectral expansion in terms of orthogonal basis in $\mathcal{H}_2(\mathcal{F})$.
Hence, $Y_i(t;\omega)$ can be spectrally represented using $P$ terms as \cite{XiuJCP03}
\begin{equation}
Y_i(t;\omega) = \sum_{p=1}^{P} Y^i_p(t) \psi_p(\omega),
\label{pc_exp}
\end{equation}
where $\psi_p(\omega)$ are orthogonal basis in $\mathcal{H}_2\left({\mathcal{F}}\right)$.
Differentiating (\ref{pc_exp}) with respect to $t$ gives
\begin{equation}
\frac{d Y_i(t;\omega)}{dt} = \sum_{p=1}^{P} \frac{d Y^i_p(t)}{dt} \psi_p(\omega),
\label{bi_eq2_1}
\end{equation}
which on Galerkin projection provide
\begin{equation}
\frac{d Y^i_p(t)}{dt} = \frac{\left\langle \frac{d Y_i(t;\omega)}{dt} , \psi_p(\omega)\right\rangle_{\Omega}}{\left\langle \psi^2_p(\omega) \right\rangle_{\Omega}}.
\label{bi_eq2_3}
\end{equation}
Having (\ref{bi_eq2_1}) and (\ref{bi_eq2_3}) in (\ref{gov_eq2}) results in (\ref{eyp_ev}).

\end{proof}

\subsection{Initial and Boundary Conditions for DBFE}
The initial conditions for DBFE are defined by specifying the mean $\overline{u}(x,0)$, the eigenfunctions $u_i(x,0)$ and the coefficients $Y^i_p(0)$, which are given by considering the bi-orthogonal expansion of the initial condition of SPDE (\ref{spde}) as
\begin{equation}
u(x,0;\omega) = \overline{u}(x,0) + \sum^{N}_{i=1} \sum^{P}_{p=1} Y^i_p(0) u_i(x,0) \psi_p(\omega).
\end{equation}
Applying the expectation operator, initial condition for the mean field is given by
\begin{equation}
\overline{u}(x,0) = E^{\omega} \left[ u(x,0;\omega) \right].
\label{mean_ini}
\end{equation}

The initial condition for eigenfunctions, $u_i(x,0)$, is given by solution of the Fredholms' integral equation of the second kind~\cite{fie_09}
\begin{equation}
\int_{X} C_{u(x_1,0),u(x_2,0)} u_i(x_1,0) d x_1 = \lambda^2_i u_i(x_2,0)
\label{fie}
\end{equation}
where $C_{u(x_1,0),u(x_2,0)}$ is the covariance function of $u(x,0;\omega)$ and $\lambda_i$ are the eigenvalues.
A Galerkin projection based method is used to solve (\ref{fie}) numerically~\cite{Huang_ijnme01}.
For a Gaussian process, all the coefficients $Y^i_p(0)$ are zero except
\begin{equation}
Y^i_2(0)=\sqrt{\lambda_i},
\label{ini_yi_norm}
\end{equation}
thus defining $Y_i(0;\omega)$ as normal variables with variance $\lambda_i$.
For a generic non-Gaussian case, the initial expansion coefficients $Y^i_p(0)$ are given by
\begin{equation}
Y^i_p(0) = \frac{\Big\langle \left\langle u(x,0;\omega) - \overline{u}(x,0),u_i(x,0) \right\rangle_{\mathcal{X}},\psi_p(\omega) \Big\rangle_{\Omega}}{\left\langle \psi^2_p(\omega) \right\rangle_{\Omega}}.
\label{yi_ini}
\end{equation}

Boundary conditions for DBFE are given by using the generic KL expansion of $h(\beta,t;\omega)$, which specifies the boundary conditions for the SPDE in (\ref{spde}), as
\begin{equation}
h(\beta,t;\omega) = \overline{h}(\beta,t) + \sum^{N}_{i=1} Y_i(t;\omega) u_i(\beta,t).
\label{bound_kl_bi}
\end{equation}
Applying the expectation operator on (\ref{bound_kl_bi}), boundary condition for the mean field is given by
\begin{equation}
\mathcal{B}\left[\overline{u}(\beta,t;\omega) \right]_{\beta \in \partial \mathcal{X}}  = \overline{h}(\beta,t) .
\label{bound_mean}
\end{equation}
Multiply (\ref{bound_mean}) by $Y_j(t;\omega)$ and apply the expectation operator
\begin{equation}
E^{\omega}\left[h(\beta,t;\omega) Y_j(t;\omega) \right] = \sum^{N}_{i=1} C_{Y_i(t) Y_j(t)} u_i(\beta,t),
\end{equation}
which can be specified in a matrix form as
\begin{equation}
{\bf u} = \Sigma^{-1} {\bf E},
\end{equation}
where $\Sigma$ is the covariance matrix.

\section{Resolution of the Gibbs Phenomena using Dynamically Bi-orthogonal Field Equations }
Key challenges involved in the numerical implementation of the DBFE for SPDEs with discontinuous solutions are: (a) numerical evaluation of the mean $\overline{u}(x,t)$ and the eigenfunctions (using (\ref{mean_ev}) and (\ref{efun_ev})) results in the development of spurious oscillations as discontinuities evolve in the solution; (b) the resultant bi-orthogonal expansion (\ref{kl_expn_dbfe}) exhibits the Gibbs phenomenon characterized by the oscillations near the discontinuity location and the slow convergence away from the discontinuity location.
This paper proposes a two step approach for the application of DBFE to the stochastic systems with discontinuous solution, that exploits the orthogonal decomposition of the spatial field in the DBFE approach.
In the first step, extension of the existing schemes is proposed to derive the non-oscillatory numerical scheme for DBFE.
In the second step, a Gegenbauer reprojection based method is proposed to mitigate the effects of the Gibbs phenomenon.
In this section, the proposed approach is described in detail.

\subsection{Numerical Scheme for Non-oscillatory Solution}
Due to discontinuities, numerical solution of (\ref{mean_ev})--(\ref{eyp_ev}) is expected to contain spurious oscillations with reduced accuracy in $\overline{u}(x,t)$ and $u_i(x,t)$.
Total Variation Diminishing (TVD) finite difference schemes are widely used in the literature to obtain non-oscillatory accurate weak solutions to hyperbolic partial differential equations~\cite{LeVeque}.
This paper demonstrates extension of a TVD scheme for numerical solution of (\ref{mean_ev})--(\ref{eyp_ev}), however, note that other numerical schemes can similarly be extended for DBFE.

For notational convenience, this subsection defines $u^t_j(\omega)$ as an approximate value of $u(x=x_j,t;\omega)$ at the grid point $x_j=j\Delta x$ and present time $t$.
Consider a numerical scheme applied to SPDE (\ref{spde}) for a given $\omega$ as
\begin{equation}
\frac{\partial u^t_j(\omega)}{\partial t} = - \frac{1}{\Delta x}\left(\hat{f}_{j+\frac{1}{2}}(\omega) - \hat{f}_{j-\frac{1}{2}}(\omega) \right),
\label{tvd_1}
\end{equation}
where $\hat{f}_{j+\frac{1}{2}}(\omega)$ is an interpolated flux inside a cell $[x_j, x_{j+1}]$.
A $(2k +1)$ point numerical scheme is defined by using $k$ points on either side of $x_j$ for interpolation, thus,
\begin{equation}
\hat{f}_{j+\frac{1}{2}}(\omega) = f(u^t_{j-k+1}(\omega),...,u^t_{j+k}(\omega)),
\end{equation}
such that \cite{LeVeque}
\begin{equation}
\hat{f}(u,...,u) = f(u).
\end{equation}
$\hat{f}_{j-\frac{1}{2}}(\omega)$ is also defined in a similar manner.
The TVD scheme (\ref{tvd_1}) can be extended to DBFE by using
\begin{equation}
\mathcal{L} \left[u(x,t;\omega);\omega \right] = - \frac{1}{\Delta x}\left(\hat{f}_{j+\frac{1}{2}}(\omega) - \hat{f}_{j-\frac{1}{2}}(\omega)\right)
\end{equation}
for numerical solution of (\ref{mean_ev})--(\ref{eyp_ev}).
The resultant numerical scheme retain non-oscillatory property in $\overline{u}(x,t)$ and $u_i(x,t)$.

In the present paper, the DBFE method is implemented using the first order central scheme proposed by Kurganov and Tadmor~\cite{kt_jpc_2000} (central KT scheme), for which numerical flux is defined as
\begin{equation}
\hat{f}_{j+\frac{1}{2}}(\omega) = -\frac{1}{2}\left[f(u^t_{j+1}(\omega))+f(u^t_{j}(\omega))\right] + \frac{1}{2}\left[a_{j+\frac{1}{2}}(\omega)\left(u^t_{j+1}(\omega)-u^t_j(\omega) \right)\right],
\label{num_flux}
\end{equation}
where $a_{j+\frac{1}{2}}(\omega)$ is the maximum local speed at $x_j$.
The gPC expansion of $a_{j+\frac{1}{2}}(\omega)$ is given by
\begin{equation}
a_{j+\frac{1}{2}}(\omega) = \sum^{P}_{p=1} \hat{a}^{j+\frac{1}{2}}_p \psi_p(\omega),
\label{gpc_spec_rad}
\end{equation}
where $\hat{a}^{j+\frac{1}{2}}_p$ are gPC expansion coefficients.
Use (\ref{gpc_spec_rad}) in numerical flux (\ref{num_flux}) to obtain
\begin{eqnarray}
E^{\omega}\left[\hat{f}_{j+\frac{1}{2}}(\omega)\right] & = & -\frac{1}{2}\big\{ E^{\omega}\left[f(u^t_{j+1}(\omega))\right] + E^{\omega}\left[f(u^t_j(\omega))\right]\big\}  \nonumber \\
~ & ~ & + \frac{1}{2}\big\{\hat{a}^{j+\frac{1}{2}}_1 [\overline{u}(x_{j+1},t)-\overline{u}(x_{j},t)] + \sum^{N}_{i=1} M^{i,j+\frac{1}{2}}_1 [u_i(x_{j+1},t) - u_i(x_j,t)] \big\} \nonumber \\
E^{\omega}\left[\hat{f}_{j+\frac{1}{2}}(\omega) Y_k(t;\omega)\right] & = & -\frac{1}{2}\big\{E^{\omega}\left[f(u^t_{j+1}(\omega)) Y_k(t;\omega) \right] + E^{\omega}\left[f(u^t_{j}(\omega)) Y_k(t;\omega) \right]\big\} \nonumber \\
~ & ~ & + \frac{1}{2}\big\{(M^{k,j+\frac{1}{2}}_1 [\overline{u}(x_{j+1},t)-\overline{u}(x_{j},t)] \nonumber \\
~ & ~ & + \sum^{N}_{i=1} T^{i,k,j+\frac{1}{2}}_1 [u_i(x_{j+1},t) - u_i(x_j,t)] \big\} \nonumber \\
\hat{f}_{j+\frac{1}{2}}(\omega) - E^{\omega}\left[\hat{f}_{j+\frac{1}{2}}(\omega)\right] & = & -\frac{1}{2} \big\{f(u^t_{j+1}(\omega))-E^{\omega}\left[f(u^t_{j+1}(\omega))\right] + f(u^t_{j}(\omega))-E^{\omega}\left[f(u^t_{j}(\omega))\right]\big\} \nonumber \\
~ & ~ & + \frac{1}{2} \big\{\sum^P_{p=2} \hat{a}^{j+\frac{1}{2}}_p [\overline{u}(x_{j+1},t)-\overline{u}(x_{j},t)] \nonumber \\
~ & ~ & + \sum^{N}_{i=1} \sum^{P}_{p=2} M^{i,j+\frac{1}{2}}_p [u_i(x_{j+1},t) - u_i(x_j,t)] \big\},
\label{num_flux_dbfe}
\end{eqnarray}
where $M_{i,j+\frac{1}{2}}= Y_i(t;\omega) a_{j+\frac{1}{2}}(\omega)$ and $T_{i,k,j+\frac{1}{2}}=Y_i(t;\omega) Y_k(t;\omega) a_{j+\frac{1}{2}}(\omega) $.

The numerical flux (\ref{num_flux_dbfe}) is used in (\ref{tvd_1}) to obtain the first order central difference scheme for numerical solution of (\ref{mean_ev})-(\ref{eyp_ev}).
$4^{th}$ order Runge-Kutta method is used for time integration in (\ref{mean_ev}) and (\ref{efun_ev}), while Euler method is used to solve ODE (\ref{eyp_ev}).
Gauss-Legendre quadrature is used to calculate inner products in spatial dimensions, while, Monte Carlo integration is used for inner products in stochastic dimensions.

\subsection{Post-processing using Gegenbauer Polynomial}
Let the solution of the SPDE in (\ref{spde}) is obtained using the DBFE method till time $t=T$ such that the discontinuities are developed. For a given $\omega$, let $[a(\omega),b(\omega)]$ denote an interval of analyticity for $u(x,T;\omega)$, where a random variable $\zeta(x;\omega)$ can be defined such that $-1\leq \zeta(x;\omega) \leq 1$.
Gibbs phenomenon can be resolved in the interval $[a(\omega),b(\omega)]$ by reprojecting the partial sum (\ref{kl_expn_dbfe}) on a suitable Gibbs complementary polynomial basis~\cite{gottlieb_11}.
In the present paper, the partial sum is reprojected on a Gegenbauer Polynomial, which is defined for a $\zeta(x;\omega)$ as follows.
\newtheorem{gegenpoly}[inpdef]{Definition}
\begin{gegenpoly}~\cite{gottlieb_11}
The Gegenbauer polynomial $C^{\lambda}_n(\zeta(x;\omega))$ is a polynomial of degree $n$ which is orthogonal over $[-1,1]$ with weight function $(1-\zeta^2(x;\omega))^{\lambda-\frac{1}{2}}$ for $\lambda \geq 0$.
The orthogonality is given by
\begin{equation}
\int^{1}_{-1} (1-\zeta^2(x;\omega))^{\lambda-\frac{1}{2}} C^{\lambda}_n(\zeta(x;\omega)) C^{\lambda}_m(\zeta(x;\omega)) d\zeta = h^{\lambda}_n \delta_{m,n},
\end{equation}
where
\begin{equation}
h^{\lambda}_n = \frac{\sqrt{\pi} C^{\lambda}_n(1) \Gamma(\lambda+\frac{1}{2})}{\Gamma(\lambda) (m+\lambda) },
\end{equation}
with
\begin{equation}
C^{\lambda}_n(1) = \frac{\Gamma(n+2 \lambda)}{m!\Gamma(2 \lambda)}.
\end{equation}
$C^\lambda_n(\zeta(x;\omega))$ can be estimated using a recurrence relationship
\begin{equation}
(n+1)C^\lambda_{n+1}(\zeta(x;\omega)) = 2 (\lambda + n) \zeta(x;\omega) C^\lambda_{n}(\zeta(x;\omega)) - (2\lambda + n -1) C^\lambda_{n-1}(\zeta(x;\omega)).
\label{recur_gegen}
\end{equation}
\end{gegenpoly}

Using orthogonality of Gegenbauer polynomial, $u(x,T;\omega)$ can be represented in terms of exponentially convergent Gegenbauer expansion series truncated at $M$ terms as
\begin{equation}
u(x,T;\omega) = \sum^{M}_{l=1} \hat{u}_l(\omega) C^{\lambda}_l(\zeta(x;\omega)),
\label{gegen_exp}
\end{equation}
where $\zeta(x;\omega)=\frac{x-\delta(\omega)}{\epsilon(\omega)}$; $\epsilon(\omega)=\frac{b(\omega)-a(\omega)}{2}$ and $\delta(\omega)=\frac{b(\omega)+a(\omega)}{2}$.
Coefficients $\hat{u}_l(\omega)$ are given by
\begin{equation}
\hat{u}_l(\omega) = \frac{1}{h^\lambda_l} \int^{1}_{-1} (1-\zeta^2(x;\omega))^{\lambda-\frac{1}{2}} C^\lambda_l(\zeta(x;\omega)) u(x(\zeta),T;\omega) d\zeta.
\label{gegen_coef_u}
\end{equation}
However, $u(x(\zeta),T;\omega)$ is not available for calculation of $\hat{u}_l(\omega)$.
Using bi-orthogonal expansion (\ref{kl_expn_dbfe}) in (\ref{gegen_coef_u}),~ approximate Gegenbauer expansion coefficients $\hat{g}_l(\omega)$ are given by
\begin{equation}
\begin{split}
\hat{g}_l(\omega) = & \frac{1}{h^\lambda_l} \Bigg[\int^{1}_{-1} (1-\zeta^2(x;\omega))^{\lambda-\frac{1}{2}} C^\lambda_l(\zeta(x;\omega)) \overline{u}(x,T) d\zeta \\ & \qquad +  \sum^{N}_{i=1} \sum^{P}_{p=1} \hat{Y}^i_p(T)\psi_p(\omega) \int^{1}_{-1} (1-\zeta^2(x;\omega))^{\lambda-\frac{1}{2}} C^\lambda_l(\zeta(x;\omega)) u_i(x,T) d\zeta \Bigg].
\end{split}
\label{gegen_coef_g}
\end{equation}
Note that (\ref{gegen_coef_g}) uses a bi-orthogonal expansion of $u(x,T;\omega)$, thus, $\hat{g}_l(\omega)$ is an approximation of $\hat{u}_l(\omega)$.
$\hat{g}_l(\omega)$ can be expanded in gPC basis as
\begin{equation}
\hat{g}_l(\omega) = \sum^{P}_{p=1} \hat{g}^l_p \psi_p(\omega),
\end{equation}
where
\begin{equation}
\hat{g}^l_p = \frac{\left\langle \hat{g}_l(\omega), \psi_p(\omega)\right\rangle_{\Omega}}{\left\langle \psi^2_p(\omega) \right\rangle_{\Omega}}.
\label{gp_coef}
\end{equation}
Coefficients $\hat{g}^l_p$ are calculated using Gaussian quadrature by evaluating (\ref{gegen_coef_g}) at collocation points, while, the integrals in (\ref{gegen_coef_g}) are calculated using Gauss-Gegenbauer quadrature.
Using $\hat{g}^l_p$, $u(x,T;\omega)$ is approximated as
\begin{equation}
u(x,T;\omega) = \sum^{M}_{l=1} \sum^{P}_{p=1} \hat{g}^l_p \psi_p(\omega) C^\lambda_l(\zeta(x;\omega)).
\label{gegen_part_sum_pc}
\end{equation}
The partial sum (\ref{gegen_part_sum_pc}) approximates $u(x,T;\omega)$ with exponential accuracy in the subinterval $[a(\omega),b(\omega)]$.
Note that determination of domain $[a(\omega),b(\omega)]$ requires location of discontinuity.
An edge detection method~\cite{gelb99,gelb01} can be used to locate points of discontinuities.

\section{Numerical Example: 1D Burgers Equation}
Efficacy of the proposed method is investigated for solution of a stochastic one-dimensional Burgers equation with uncertain input conditions.
Note that previous work has utilized deterministic Burgers equation extensively~\cite{yang_92}, while, there have been recent investigation on solution of stochastic Burgers equation~\cite{burg_jcp_09,PoetteJCP09}.
Coupled with the ease of implementation, stochastic Burgers equation provide an appropriate context to investigate the efficacy of the proposed method for mitigation of the Gibbs phenomenon.

A one dimensional inviscid stochastic Burgers equation is given by
\begin{equation}
\frac{\partial }{\partial t} \left( u(x,t;\omega) \right) = - \frac{\partial }{\partial x} \left( \frac{u^2(x,t;\omega)}{2} \right),
\label{spde_be}
\end{equation}
where $x \in [-1,1]$.
In the present paper, the stochastic Burgers equation is investigated for uncertain initial condition given by
\begin{equation}
u(x,0;\omega) = s G(x;\omega),
\label{ini_u}
\end{equation}
where $G(x;\omega)$ is a Gaussian process with mean
\begin{equation}
\overline{u}(x,0) = u_b - \displaystyle{\tan^{-1}}\left(x-x_0\right),
\label{ini_mean}
\end{equation}
where $u_b$ and $x_0$ are user defined constants, and the covariance function
\begin{equation}
C_{u(x_1,0),u(x_2,0)} = \sigma^2 \exp\left(-\lambda \mid x_1-x_2 \mid \right),
\label{ini_cov}
\end{equation}
where $\sigma$ is standard deviation and $\lambda$ is the correlation length.
In this paper results are presented for $s=0.1$.
The boundary condition is specified using
\begin{equation}
u(x,t;\omega) = u(\beta,0;\omega).
\end{equation}


\subsection{Monte Carlo Method for Stochastic One-dimensional Burgers Equation}
\begin{figure}[h!]
  \begin{center}
  \begin {tabular}{l l}
  \subfigure []{\includegraphics[width=3in, height=2.85in] {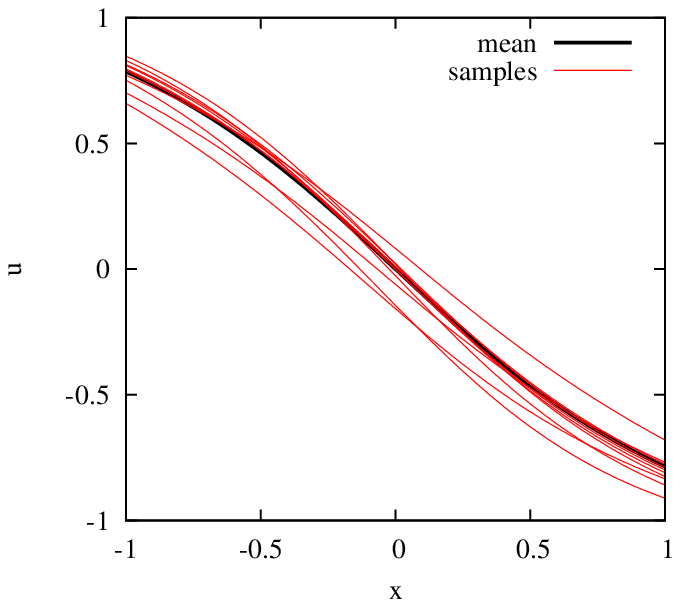}} &
   \subfigure []{\includegraphics[width=3in, height=2.85in] {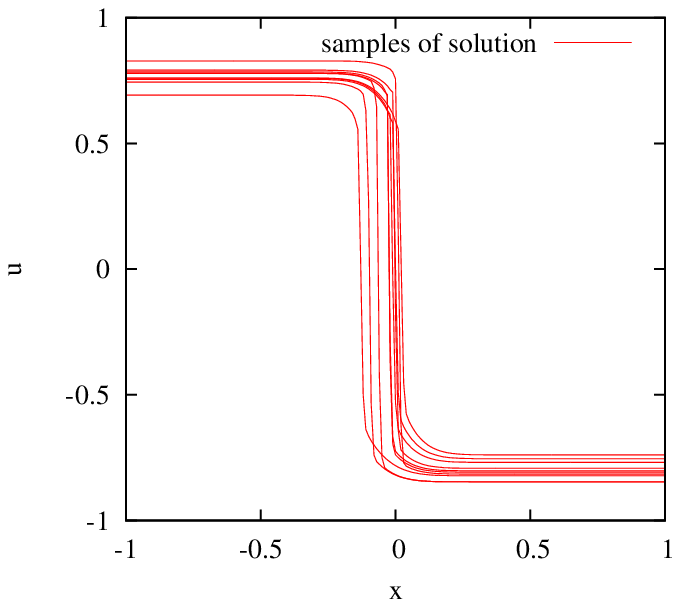}} \\ 
   \end{tabular}
  \end{center}
 \caption{Solution of SPDE (\ref{spde}) using Monte Carlo method. Figure a) shows samples of initial condition; figure b) shows solution of samples corresponding to the initial conditions in a).}
\label{ini_cond_mean}
\end{figure}

To investigate accuracy of the proposed DBFE method, its numerical results are compared with the Monte Carlo simulation.  
The Monte Carlo method for a stochastic one-dimensional Burgers equation is applied by solving the deterministic Burgers equation with initial condition defined using samples of $u(x,0;\omega)$.   
In this section, results are presented for a test case with $u_b = 0, x_0 = 0$.
For each sample, explicit fourth order Runge-Kutta scheme is used for time evolution while the first order central KT scheme~\cite{kt_jpc_2000} (see equation (\ref{num_flux})) is used in spatial dimension.
The numerical solutions are obtained for $\Delta x = 0.01$ and $\Delta t = 10^{-4}$.
Total 10000 samples are collected for the Monte Carlo simulation.  
Figure \ref{ini_cond_mean}(a) shows several realizations of initial conditions, where the mean is shown with emphasis, while respective solutions at time $t=1.1s$ are shown in Figure \ref{ini_cond_mean} (b).
Shock formation is observed for all the samples, however, shock location varies  depending on the initial condition.
\subsection{Application Results}
Use the bi-orthogonal expansion of $u(x,t;\omega)$ in (\ref{spde_be}) to obtain
\begin{eqnarray}
\mathcal{L}\left[u(x,t;\omega);\omega \right] & = & -\frac{\partial }{\partial x}  \frac{\overline{u}^2(x,t)}{2} - \sum^{N}_{i=1} \sum^{P}_{p=1} Y^i_p(t) \psi_p(\omega) \frac{\partial }{\partial x} \left(\overline{u}(x,t) u_i(x,t)\right) \nonumber \\
~ & ~ & - \sum^{N}_{i=1} \sum^{N}_{j=1} \sum^{P}_{p=1} \sum^{P}_{q=1} Y^i_p(t) Y^j_q(t) \psi_p(\omega) \psi_q(\omega) \frac{\partial }{\partial x} \left(\frac{u_i(x,t) u_j(x,t)}{2}\right).
\label{f_be}
\end{eqnarray}
The differential operator (\ref{f_be}) is used in (\ref{num_flux_dbfe}) to derive the DBFE governing equations (\ref{mean_ev}-\ref{eyp_ev}).

\begin{figure}[h!]
  \begin{center}
  \begin {tabular}{l l}
  \subfigure []{\includegraphics[width=3in, height=2.85in] {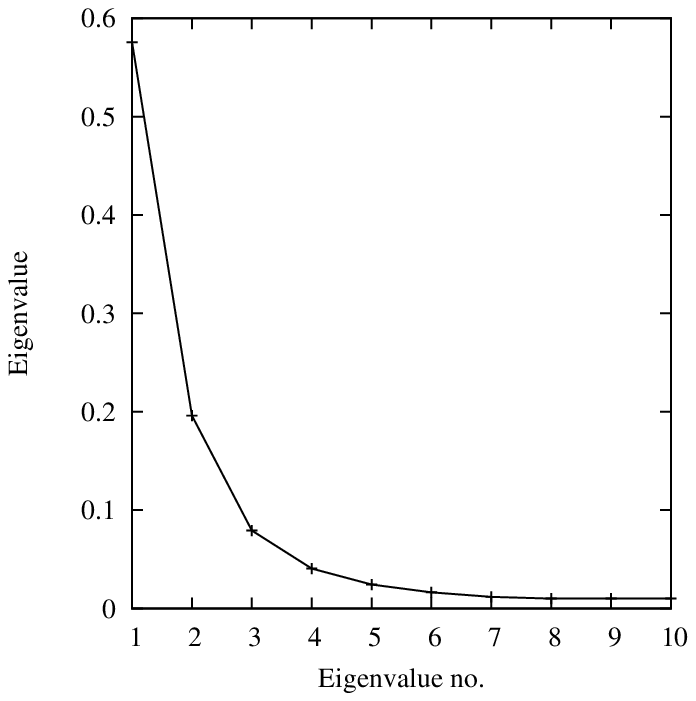}} &
    \subfigure []{\includegraphics[width=3in, height=2.85in] {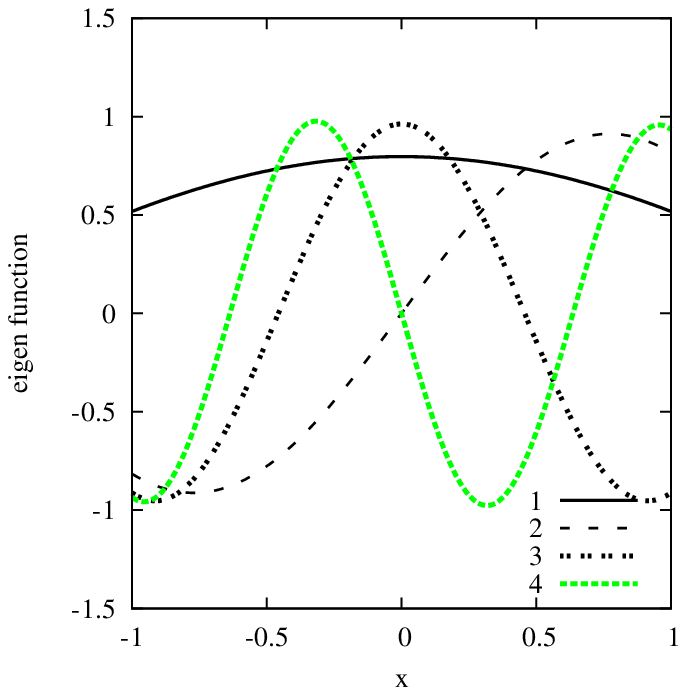}} \\
   \end{tabular}
  \end{center}
 \caption{Eigenfield for uncertain initial condition. a) eigenvalues and b) first four eigenfunctions }
\label{ini_eigen}
\end{figure}

Initial condition for the mean $\overline{u}(x,t)$ is specified using (\ref{ini_mean}), while the initial condition for $u_i(x,t)$ and $Y^i_p(t)$ are given by solution of the eigenvalue problem for the covariance function (\ref{ini_cov}). 
Note that the eigenvalues for the covariance function (\ref{ini_cov}) decreases rapidly with only first three eigenmodes dominant, thus, $N=3$ suffices for the accurate approximation of the uncertain initial condition.
However, to account for a highly nonlinear nature of the Burgers equation, higher value of $N$ may be required.
Figure \ref{ini_eigen} shows eigenvalues and first four eigenfunctions of the uncertain initial condition with $\sigma=0.5$ and $\lambda=1.0$.

\subsubsection{Non-oscillatory Solution using the Proposed Numerical Scheme}
Efficacy of the numerical scheme described in the section 3.1 to obtain non-oscillatory solutions for mean and eigenfunctions is investigated by implementing the DBFE method using the first order central KT scheme (\ref{num_flux_dbfe}), central difference scheme (terms involving $\hat{a}^{j+\frac{1}{2}}_p$, $M^{i,j+\frac{1}{2}}_p$ and $T^{i,k,j+\frac{1}{2}}_p$ are neglected in (\ref{num_flux_dbfe}))  and the first order central KT scheme in mean (terms involving $\hat{a}^{j+\frac{1}{2}}_1$ and $M^{i,j+\frac{1}{2}}_1$ are retained only for the first term in (\ref{num_flux_dbfe})).

\begin{figure}[h!]
  \begin{center}
  \begin {tabular}{l l}
  \subfigure []{\includegraphics[width=3in, height=2.85in] {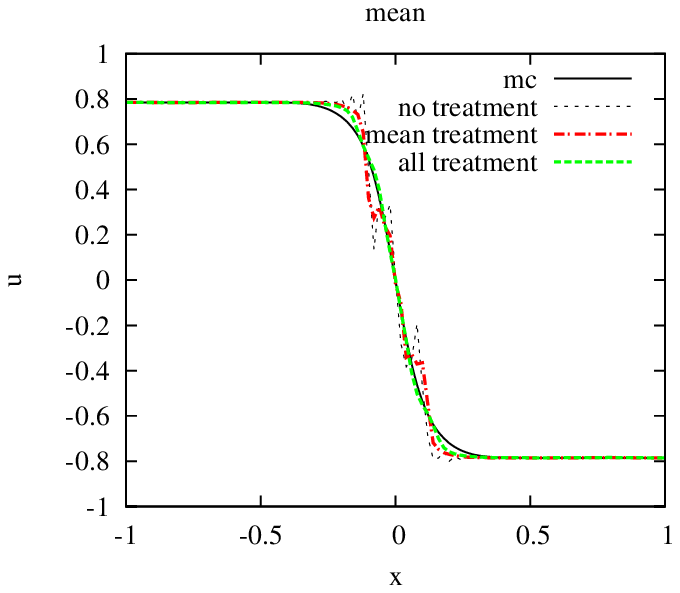}} &
    \subfigure []{\includegraphics[width=3in, height=2.85in] {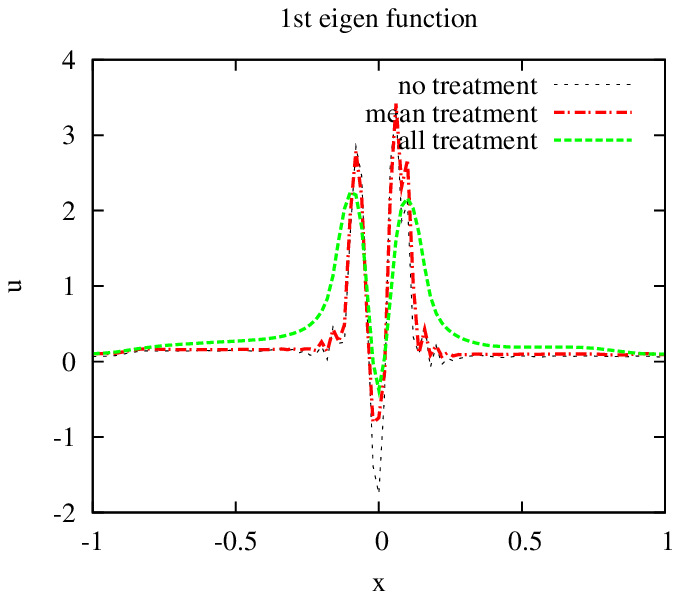}} \\
   \end{tabular}
  \end{center}
 \caption{Figure shows comparison of numerical solution for a) mean and b) first eigenfunction, obtained by implementing the proposed DBFE method using the central difference scheme, first order KT scheme in mean and full first order central KT scheme. The results are obtained using the first three eigenmodes (N=3) in bi-orthogonal expansion.}
\label{comp_gibbs}
\end{figure}

Figure \ref{comp_gibbs} shows comparison of results for (a) mean and (b) first eigenfunction for $N=3$.
With central difference scheme, oscillations are observed in both mean and eigenfield.
When central KT scheme is used only in mean field and central difference for eigenfield, oscillations in mean are comparatively reduced.
However oscillations are still present in eigenfield, as can be observed for first eigenfunction.
When full central KT scheme is used, oscillations in both mean and eigenfield are resolved.
Note that only averaged statistics in terms of expected values and covariance functions affect time evolution of mean and eigenfields, hence, no special treatment is necessary for random field.

\subsubsection{Characteristics of DBFE Solutions}
The characteristics of the numerical solution for DBFE before post-processing is investigated.
In order to investigate effect of truncation of KL expansion, Burgers equation is solved for $N=3$, $N=5$ and $N=7$.
To compare results with Monte Carlo simulations, $u(x,t;\omega)$ is reconstructed for every sample using (\ref{kl_expn_dbfe}).
Accuracy of the DBFE method is compared with Monte Carlo using relative error in $L_1(\Omega)$ norm, which is given by
\begin{equation}
e^{L_1} \left(x,t\right) = \frac{\int_{\Omega} \int_{\delta x} \left| u(x,t,\omega) - \left( \overline{u}_b (x,t) + \sum^{N}_{i=1} \sum^{P}_{p=1} Y^i_p(t) u_i(x,t) \psi_p(\omega) \right) \right| dx d\mathcal{P}(\omega)}{\int_{\Omega} \int_{\delta x} \mid u(x,t,\omega) \mid dx d\mathcal{P}(\omega)}.
\label{l1_error}
\end{equation}

\newtheorem{rmklone}{Remark}
\begin{rmklone}
Note that the $L_1$-error in (\ref{l1_error}) is defined for a small domain $\delta x$, and thus is not a point error.
In the present paper, the $L_1$-error at a point is defined using the adjoining two cells.
\end{rmklone}

\begin{figure}[h!]
  \begin{center}
  \includegraphics[width=3in, height=2.85in] {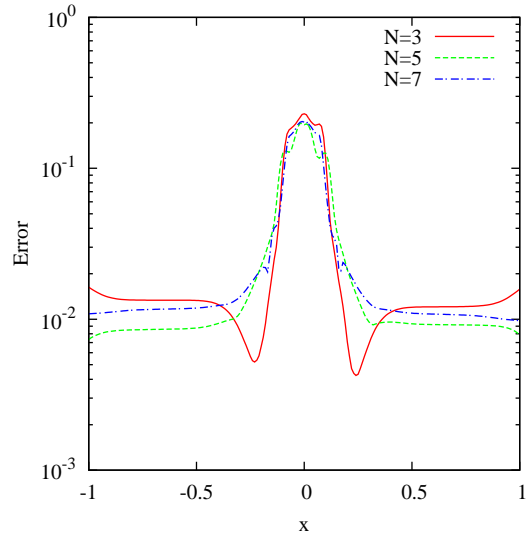}
  \end{center}
 \caption{Figure shows $L_1(\Omega)$ error for DBFE method. The error is calculated against the Monte Carlo method using 10000 samples.}
\label{comp_err}
\end{figure}

Figure \ref{comp_err} shows the $L_1$-error of the DBFE method relative to the Monte Carlo method.
$L_1(\omega)$ error of the order of $10^{-2}$ is observed away from the shock location, however, near the shock location, error of the order of $10^{-1}$ is observed that does not reduce significantly with increasing $N$.

\begin{figure}[h!]
  \begin{center}
  \begin {tabular}{l l}
  \subfigure []{\includegraphics[width=3in, height=2.85in] {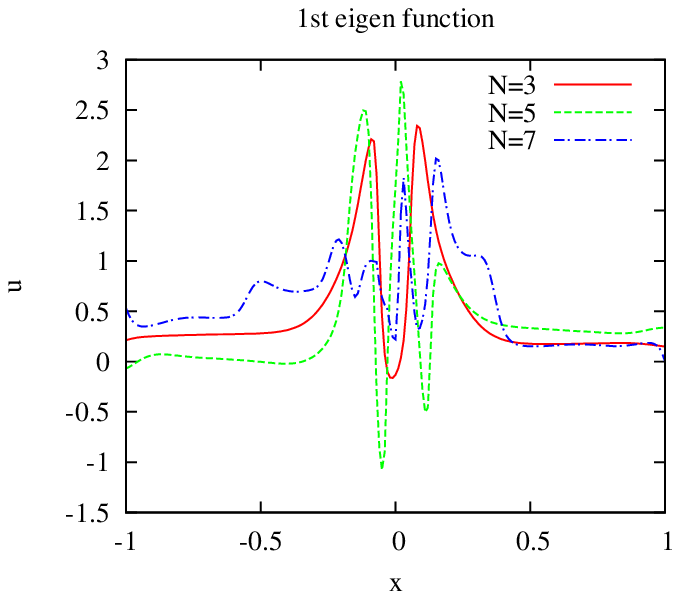}} &
    \subfigure []{\includegraphics[width=3in, height=2.85in] {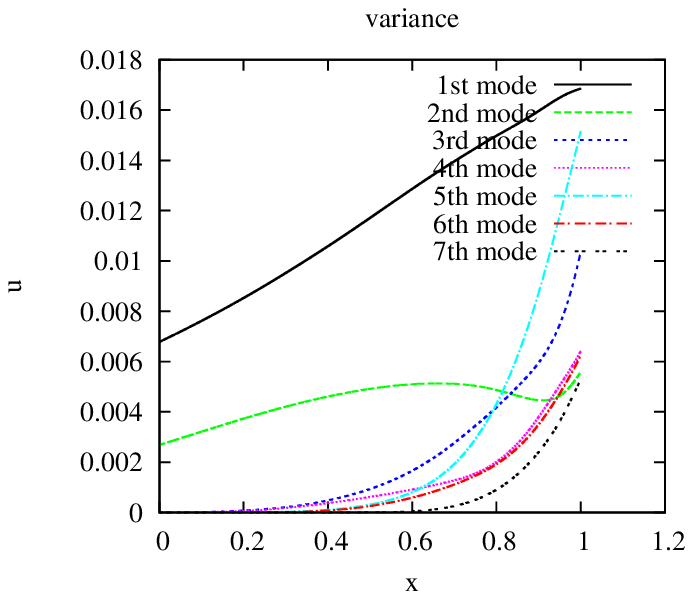}} \\
   \end{tabular}
  \end{center}
 \caption{Figure shows a) 1st eigenfunction for $N=3$, $N=5$ and $N=7$, and b) variance for all modes (N=7)}
\label{comp_eigen}
\end{figure}

Figure \ref{comp_eigen} (a) shows the first eigenfunction at $t=1.1s$ for $N=3,5,7$.
The eigenfunction shows periodic oscillations near shock, with number of modes increasing with $N$.
Figure \ref{comp_eigen} (b) shows evolution of variance of different modes over time for $N=7$.
Note that the modes with negligible variance at $t=0$ become dominant over time evolution.
This effect can be significantly observed for the $5^{th}$ eigenmode, which has negligible variance at $t=0$ but becomes the second most dominant mode over the time evolution.
Thus, it is necessary to use higher number of modes, even though the modes show negligible variance at $t=0$.

\begin{figure}[h!]
  \begin{center}
    \includegraphics[width=3in, height=2.85in] {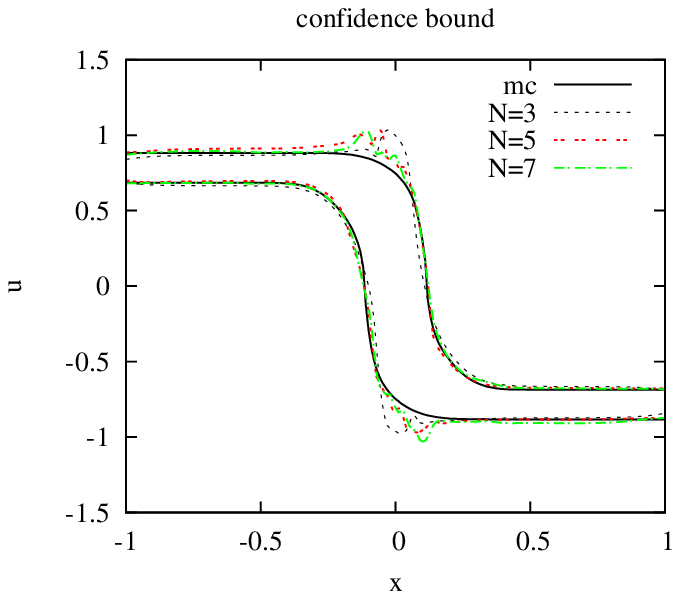} \\
  \end{center}
 \caption{Figure shows comparison of 90\% confidence bound.}
\label{comp_conf}
\end{figure}

Figure \ref{comp_conf} shows comparison of the 90\% confidence bound for the Monte Carlo and the DBEE method. 
Note that the confidence bound shows significant disagreement near shock location for all the test cases. 
The discrepancy in the confidence bound is due to the oscillations in the samples near the shock location, characterizing the Gibbs phenomenon for each sample.

The key observations from the results presented here can be summarized as: a) numerical solution of the DBFE exhibits oscillations near the discontinuity, b) $L_1$-error is maximum at the shock location, c) The amplitude of the maximum $L_1$-error does not decrease by increasing the number of eigenfunctions, $N$.
These observations are characteristics of the Gibbs phenomenon, indicating the need for appropriate post-processing.
Efficacy of the proposed post-processing to mitigate the effect of the Gibbs phenomenon is demonstrated in the following.

\subsubsection{Mitigation of Gibbs Phenomenon using Post-processing}
The Gegenbauer reprojection based approach, outlined in section 3.2, is used for post-processing to mitigate the effect of the Gibbs phenomenon.
The bi-orthogonal expansion of $u(x,t;\omega)$ is reprojected on $7^{th}$ order Gegenbauer polynomials, while total 7 coefficients are used in the expansion (\ref{gegen_part_sum_pc}).
Integrals involving the Gegenbauer polynomials are numerically evaluated using Gauss-Gegenbauer quadrature with 100 nodes, whereas, the integrals involving the polynomial chaos basis are numerically solved using Monte Carlo integration.
In the present paper, shock is assumed to be located at a point with highest slope where $u=0$ line is crossed.
However, the proposed method can be implemented using any edge detection method for shock localization.

\begin{figure}[h!]
  \begin{center}
  \includegraphics[width=6in, height=2.85in]{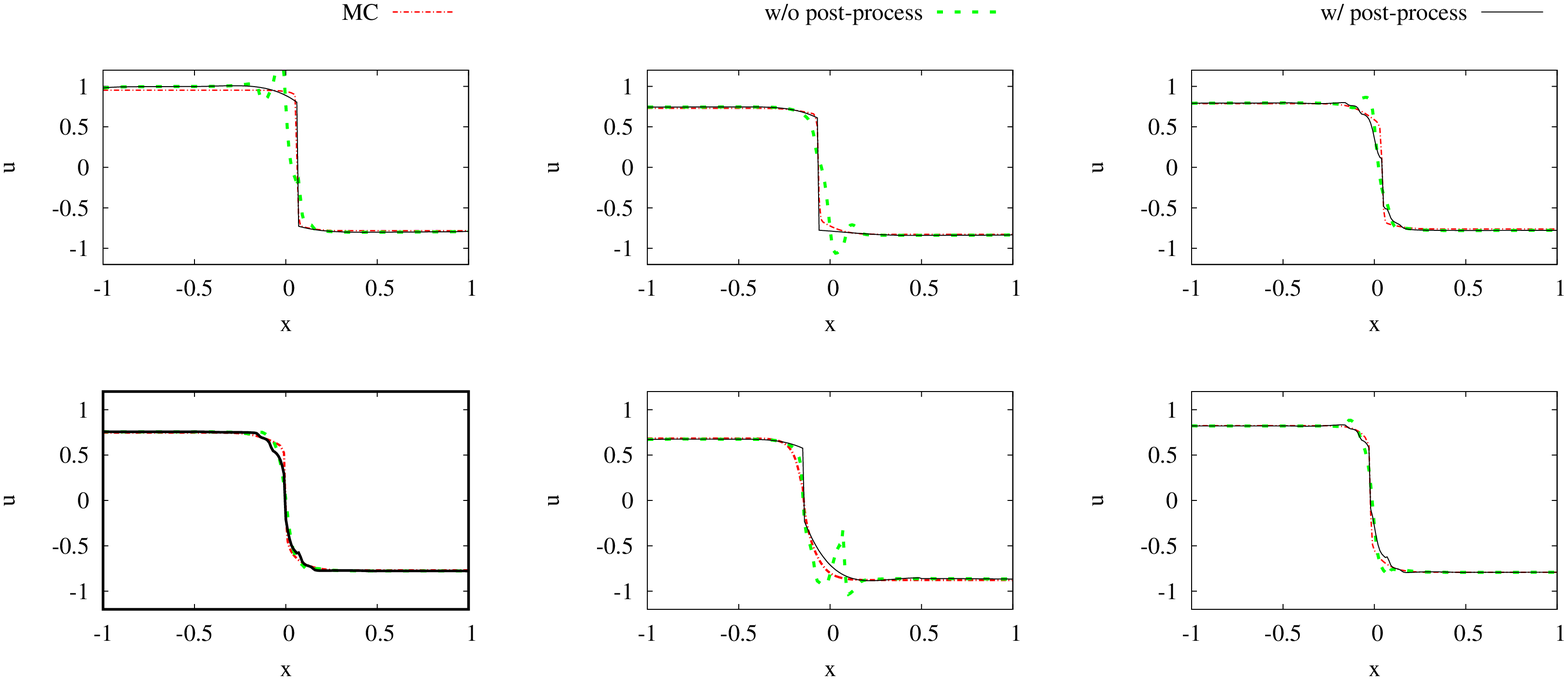}
  \end{center}
 \caption{Figure shows comparison of DBFE method with Monte Carlo samples. Comparison without and with post-processing are shown for first six samples.}
\label{post_process}
\end{figure}

Resultant samples after post-processing using reprojection method are compared with samples without post-processing in Figure \ref{post_process} (the first six samples are shown in the figure).
The respective Monte Carlo samples are also shown in the figure.
Resolution of the Gibbs phenomenon can be observed for individual samples.
For all the samples, oscillations are observed for approximations of $u(x,t;\omega)$ obtained using DBFE method.
Magnitude of oscillations is low for samples near mean and increases away from mean.
The oscillations show three distinct behaviors: (1) when shock is located at $x>0$, oscillations are observed on left side of shock location; (2) when shock is located at $x<0$, oscillations are observed on right side of the shock location; and (3) when shock is located near $x=0$, oscillations are observed on both the sides with low magnitude.

\begin{figure}[h!]
  \begin{center}
\begin {tabular}{l l}
  \subfigure []{\includegraphics[width=3in, height=2.85in] {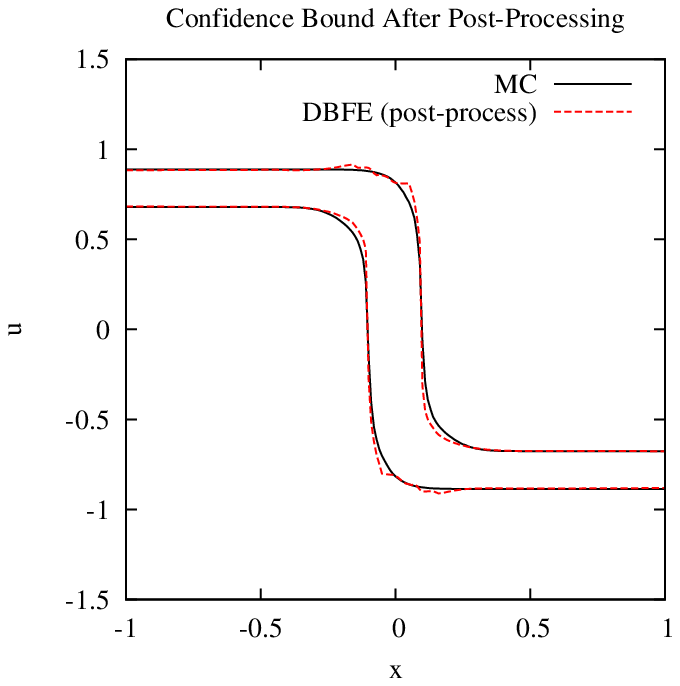}} &
    \subfigure []{\includegraphics[width=3in, height=2.85in] {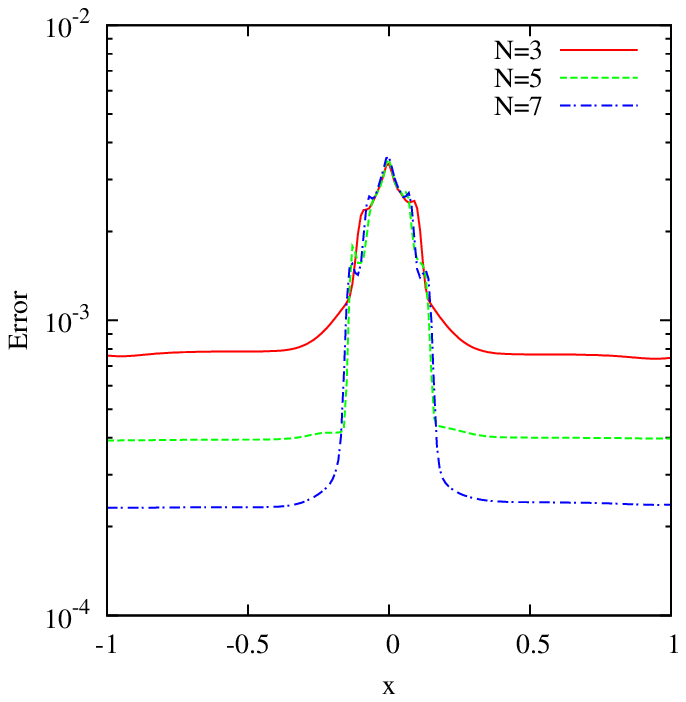}} \\
   \end{tabular}
  \end{center}
 \caption{Figure shows a) confidence bounds using post-processing b) $L_1$ error}
\label{comp_gb}
\end{figure}

Figure \ref{comp_gb} a) shows comparison of 90\% confidence bound of DBFE method after post-processing with Monte Carlo simulation.
Using post-processing, agreement with Monte Carlo simulation is significantly improved.
Figure \ref{comp_gb} b) shows $L_1$ error after post-processing.
Comparing with Figure \ref{comp_err}, it can be observed that the error is reduced at all locations by the order of $10^{-2}$.
At mean shock location, $L_1$ error is of the order of $10^{-3}$.

\begin{figure}[h!]
  \begin{center}
  \begin {tabular}{l l}
  \subfigure []{\includegraphics[width=3in, height=2.85in] {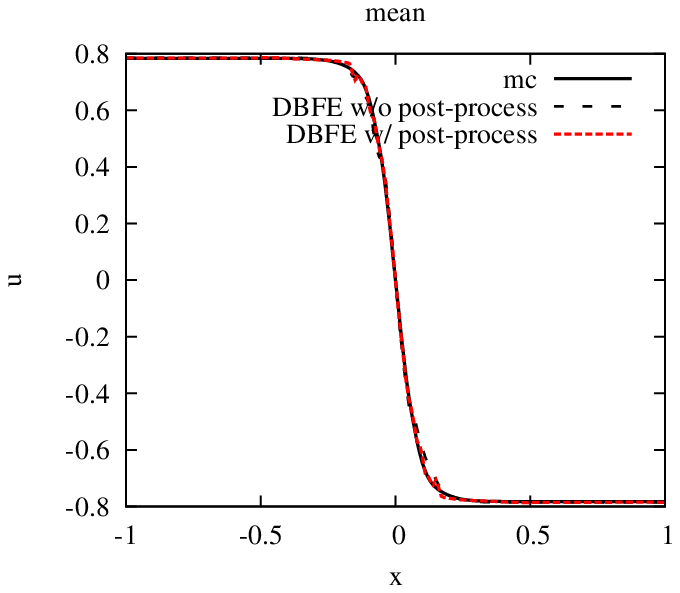}} &
    \subfigure []{\includegraphics[width=3in, height=2.85in] {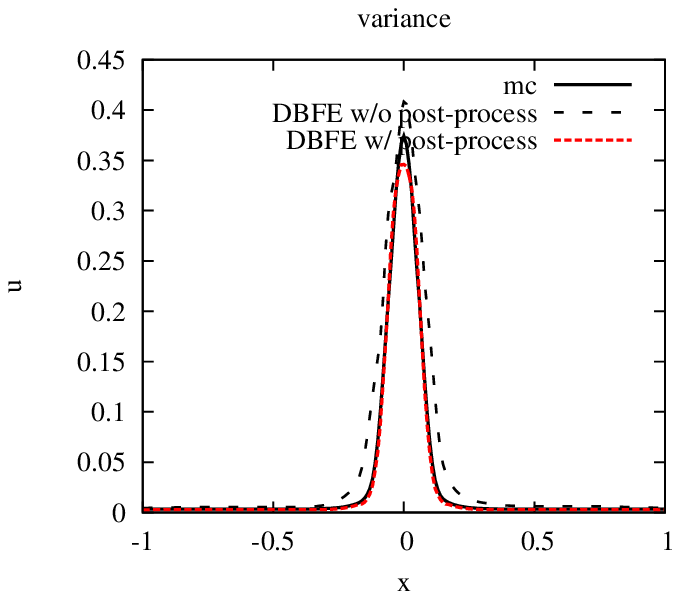}} \\
  \subfigure []{\includegraphics[width=3in, height=2.85in] {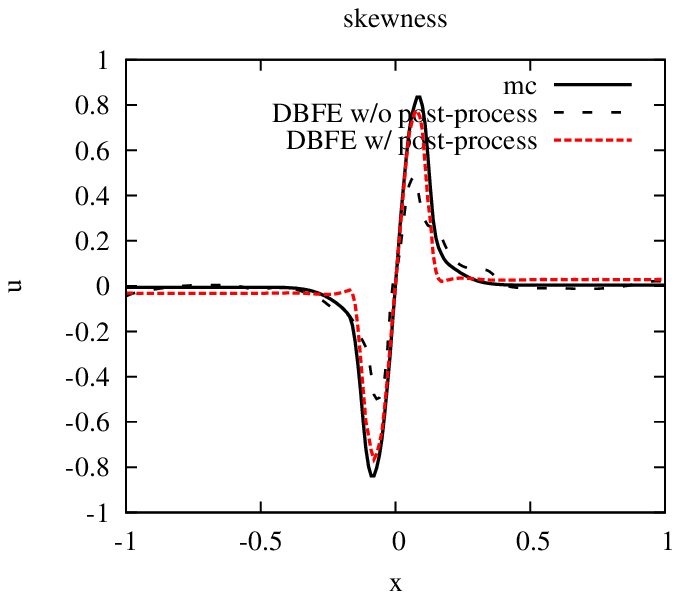}} &
    \subfigure []{\includegraphics[width=3in, height=2.85in] {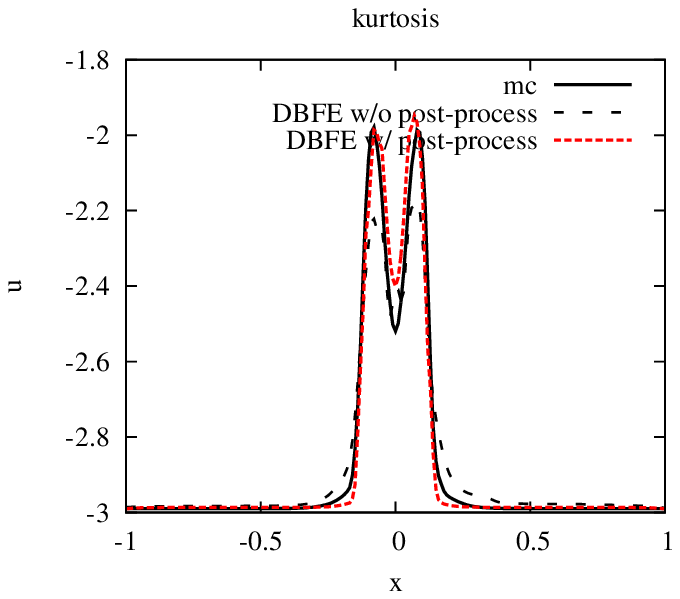}} \\
   \end{tabular}
  \end{center}
 \caption{Figure shows comparison of first four moments for Monte Carlo and DBEF. a) mean b) variance c) skewness d) kurtosis}
\label{comp_moments}
\end{figure}

Figure \ref{comp_moments} shows comparison of first four moments of DBFE solution with the Monte Carlo simulation.
The comparison is shown for both without and with post-processing cases.
For the higher moments, agreement with the Monte Carlo method improves significantly after the post-processing.
Mean of the solution obtained using the DBFE method without post-processing matches well with the Monte Carlo method (shown in Figure \ref{comp_moments} (a)).
Note that the mean is not contaminated by the Gibbs phenomenon, resulting in the good match between DBFE method without post-processing and the Monte Carlo method, while, the subsequent post-processing retains the accuracy of the mean.
Figure \ref{comp_moments} (b) shows the variance as a function of spatial location $x$.
The variance is low in the region away from the shock location, whereas, significant increase in the variance is observed at mean shock location.
Good agreement between the Monte Carlo and the DBFE method is observed in the region away from the shock location, however, nontrivial difference is observed without post-processing near the mean shock location, with DBEF predicting higher variance than the Monte Carlo.
The high variance in DBFE predictions results due to the oscillations near shock location characterizing the Gibbs phenomenon, which are mitigated using the post-processing, providing the close agreement with the Monte Carlo method.
Figure \ref{comp_moments} also shows comparison for (c) skewness and (d) kurtosis.
Skewness is zero in a region away from the shock location, indicating symmetric probability distribution.
In a region near shock, skewness is negative in the left side of the mean shock location and positive on the right, with significantly high absolute value near the mean shock location $x=0$.
Similar results are obtained for kurtosis (Figure \ref{comp_moments} (d)) where high values are obtained near shock.
It can be observed that DBFE method has predicted the trend correctly for skewness and kurtosis, while the close agreement with the Monte Carlo method is obtained after post-processing.

\subsection{Computational Efficiency of the Proposed Method}
\begin{figure}[h!]
  \begin{center}
  \includegraphics[width=3in, height=2.85in]{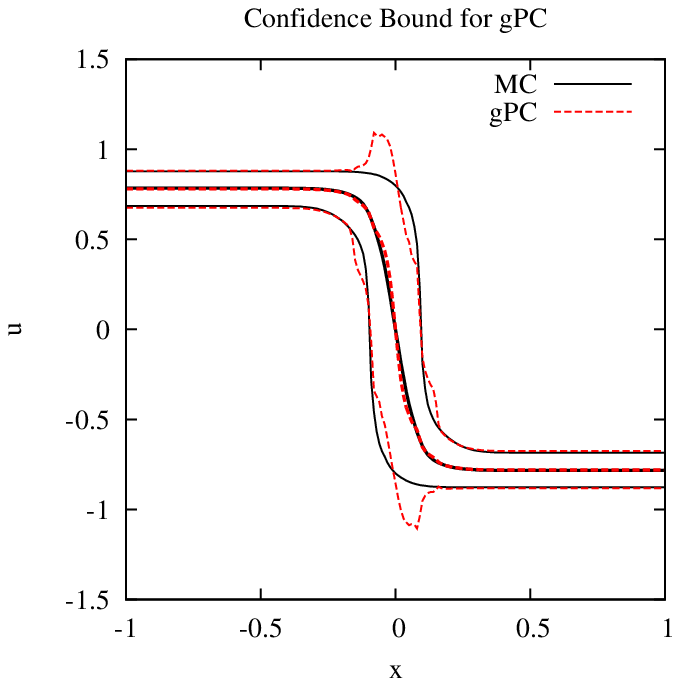}
  \end{center}
 \caption{Figure shows comparison of the confidence bound obtained using the gPC method with the Monte Carlo method. Comparison for mean is also shown in the figure.}
\label{conf_bound_gpc}
\end{figure}

The computational cost of the numerical implementation of the proposed DBFE method is compared with the Monte Carlo and the gPC method.
See Pettersson et al. \cite{burg_jcp_09} for the governing equations of the gPC method applied to stochastic Burgers equation.  
The method is implemented with $3^{rd}$ order gPC basis.
Figure \ref{conf_bound_gpc} shows the 90\% confidence bound for the gPC method, which is compared with the Monte Carlo method.
Comparison for the mean is also shown in the figure.
The mean obtained using the gPC method agrees well with the Monte Carlo method, however, the confidence bound shows oscillation near the mean shock location, demonstrating the existence of the Gibbs phenomenon for solution of the SPDE (\ref{spde}) using the gPC method.
Note that the similar observations are reported earlier in the literature for solution of the stochastic Burgers equations using the gPC method \cite{burg_jcp_09}.

\begin{figure}[h!]
  \begin{center}
  \includegraphics[width=3in, height=2.0in] {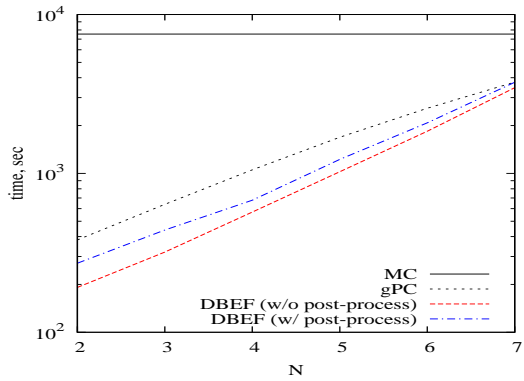}
  \end{center}
 \caption{Figure shows comparison of computational time for simulation. 10000 samples are used for the Monte Carlo method. $3^{rd}$ order Hermite polynomials are used as basis for DBFE and gPC.}
\label{comp_time}
\end{figure}

To investigate the computational efficiency of the proposed method, the Monte Carlo method, the gPC method and the DBFE are implemented on a desktop computer with Intel Core i5 CPU.
Total 10000 samples are used for the Monte Carlo method, while, $3^{rd}$ order polynomial chaos basis is used for the DBFE and the gPC method.
Time required for simulations is obtained using the FORTRAN intrinsic routine \emph{cpu\_time}.
The comparison of CPU time is shown in Figure \ref{comp_time}.
Both the gPC and the DBFE methods provide considerable computational speed-up over the Monte Carlo method, with the computational time requirement increasing with the number of eigenfunctions used.
Even for $N=7$, speed up of the order of 3 is obtained as compared to Monte Carlo method.
The computational cost of the DBFE method is lower than the gPC method, with the ratio of the CPU time for gPC and the DBFE decreasing with increasing $N$.
Although the post-processing increases the computational cost of the DBFE method, the DBFE method still remains computationally efficient with the CPU time approaching that of the gPC with $N$.
Note that the computational cost of the proposed method depends on the complexity of the differential operator $\mathcal{L}\left[u(x,t;\omega);\omega\right]$, however, the computational cost of the post-processing remains constant irrespective of the complexity of the SPDE.

\subsection{Choice of Gegenbauer Parameters} 
The proposed reprojection method requires specification of parameter $\lambda$ of $C^\lambda_l(x)$ and maximum number of expansion terms $M$.
Total error in the Gegenbauer reprojection method eminates from error due to approximation of a function using a finite Gegenbauer series expansion and roundoff error in computation of Gegenbauer polynomials $C^\lambda_n(x)$.
Roundoff error in evaluation of $C^\lambda_n(x)$ increases with increasing $\lambda$ and $n$~\cite{gelb_jsc_04}, rendering very large  $\lambda$ and $M$ unsuitable.   

\begin{figure}[h!]
  \begin{center}
  \includegraphics[width=3in, height=2.85in] {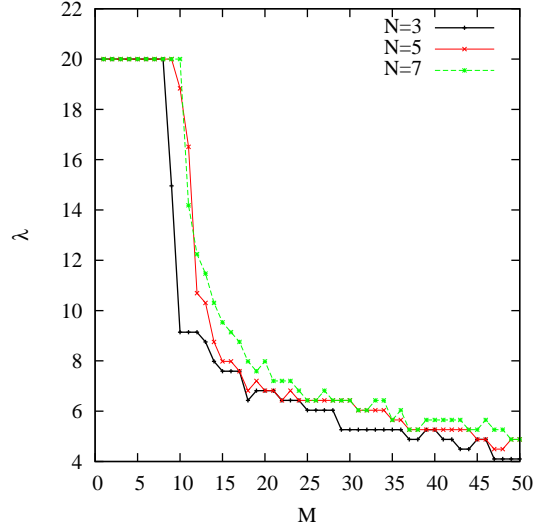}
  \end{center}
 \caption{Figure shows limiting value of $\lambda$ for given tolerance.}
\label{tol_nk}
\end{figure}

To investigate the effect of choice of parameters of Gegenbauer polynomial, proposed post-processing method is applied using different values of $\lambda$ and $M$.
Figure \ref{tol_nk} shows minimum value of $\lambda$ for which $L_1$ error is greater than the tolerance for different $N$.
For given $M$, limiting value of $\lambda$ decreases with $N$. 
In the present study, $TOL=0.1$ is used. 
note that since $u(x,t;\omega)$ is linear on the both side of shock, $\hat{g}_1(\omega)$ is dominant for all the samples, while values of higher coefficients are very low.
The higher coefficients control non-linearity near shock.
For $M=3$, highest order of Gegenbauer polynomial is quadratic in nature, which can not properly capture non-linearity near shock. 
Since the values of coefficients decrease rapidly, roundoff errors dominates for higher Gegenbauer polynomials resulting in resulting in oscillations near shock discontinuity.
For the test case presented in this paper, choice of $\lambda=7.0$ and $M=7$ provide trade-off between requirement for capturing non-linearity near shock and lower round-off error.
Although the choice of parameters is heuristic in nature, a robust optimum parameter selection method can be developed based on already established techniques for deterministic solutions~\cite{gelb_jsc_04}.

\subsection{Robustness of the Method}
\begin{figure}[h!]
  \begin{center}
  \includegraphics[width=3in, height=2.85in] {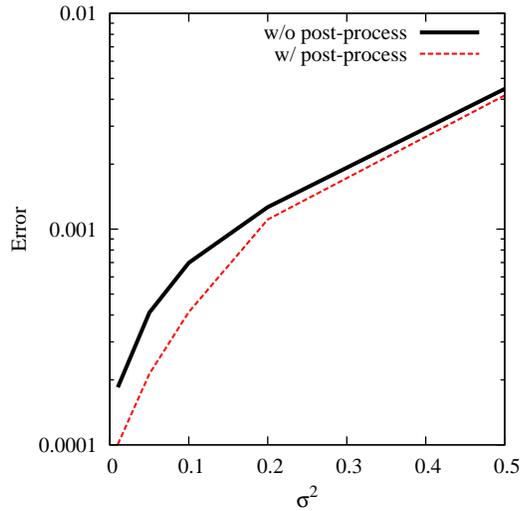}
  \end{center}
 \caption{Figure shows $L_1(\Omega)$ error as function of the variance $\sigma^2$.}
\label{comp_err_sigma}
\end{figure}

Robustness of the proposed method is investigated for following covariance functions: 1) squared exponential $C(x_1,x_2) = \sigma^2 exp\left( -(x_1 - x_2)^2 \right)$, 2) triangular $C(x_1,x_2)= (1.0-t (\mid x_1 - x_2 \mid) ) exp\left( -\mid x_1 - x_2 \mid \right)$ and 3) uniformly modulated $C(x_1,x_2)=\sigma^2 exp\left(-(x_1+x_2)\right) exp\left( -\mid x_1 - x_2 \mid \right)$.
Figure \ref{comp_err_sigma} shows $L_1$ error for different values of $\sigma^2$ when squared exponential covariance function is used to represent uncertainty in initial conditions.
$L_1$ error increases exponentially with increasing $\sigma^2$, while, the error decreases after post-processing for all $\sigma^2$.
For high $\sigma^2$, coefficients of higher modes of eigenfield become significant, necessitating the use of high $N$.
Thus, error due to truncation of spectral expansion at $N$ terms dominates the $L_1$ error for high $\sigma^2$, resulting in lesser improvement using post-processing.

\begin{figure}[h!]
  \begin{center}
  \includegraphics[width=6in, height=1.5in] {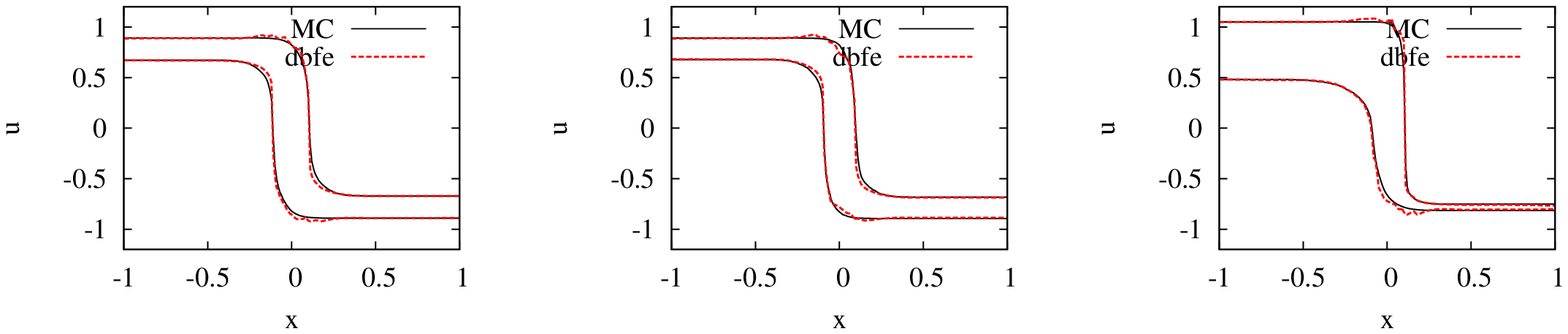}
  \end{center}
 \caption{Figure shows comparison of confidence bound for different test cases.}

\label{conf_bound_tc}
\end{figure}

Figure \ref{conf_bound_tc} shows comparison of 90\% confidence bound obtained using MC and DBFE for squared exponential, triangular and uniformly modulated covariance function.
Confidence bounds for DBFE method agrees well with MC, demonstrating robustness of the proposed method to resolve Gibbs phenomenon for different choices of covariance functions.

\section{Concluding Remarks}
This paper have proposed a dynamic bi-orthogonality based spectral projection method for uncertainty quantification for systems with discontinuous solutions.
The proposed approach is implemented in two steps: in the first step, input uncertainty is propagated to the system response using DBFE method, while, in the second step, effect of the Gibbs phenomenon is mitigated by reprojecting the mean and the eigenfield on the Gegenbauer polynomials.
Efficacy of the proposed method have been investigated for solution of a one-dimensional stochastic Burgers equation.
The numerical results presented in this paper have demonstrated the ability of the proposed post-processing method to mitigate the effect of the Gibbs phenomenon as discontinuities develop in the solution.
Note that the proposed method does not require a-priory knowledge of the shock location, thus, a generic implementation for a variety of applications can be achieved by extending any numerical scheme to the DBFE, as demonstrated in this paper.
The DBFE method is found to be computationally efficient than the gPC method, thus, the method becomes an attractive alternative to the gPC for solution of SPDEs.

\bibliographystyle{siam}
\bibliography{References_siam}
					

\end{document}